\documentclass[12pt,a4paper,twoside]{article}
  \usepackage[reqno,sumlimits]{amsmath}
  \usepackage{amsfonts,amssymb,amsthm,amsbsy,bbm,times} %macht Kopf
    \advance\textheight by \topskip
    \topmargin \paperheight \advance\topmargin by -\textheight 
    \divide\topmargin by 2 \advance\topmargin by -1in 
    \advance\topmargin by -\baselineskip
    \advance\topmargin by -\headsep
    \oddsidemargin \paperwidth \advance\oddsidemargin by -\textwidth
    \divide\oddsidemargin by 2 \advance\oddsidemargin by -1in
    \evensidemargin \oddsidemargin
\newtheorem{theorem}{Theorem}[section]
\newtheorem{proposition}{Proposition}[section]
\newtheorem{lemma}{Lemma}[section]
\newtheorem{corollary}{Corollary}[section]
\theoremstyle{definition}
\newtheorem{remark}{Remark}[section]

\makeatletter
  \@addtoreset{equation}{section}
  
\newcommand{\bel}{\begin{equation} \label}
\newcommand{\ee}{\end{equation}}

\newcommand{\ldz}{E \downarrow 0}

\newcommand{\xp}{X_\perp}
\newcommand{\bx}{{\bf x}}
\newcommand{\rt}{{\mathbb R}^{3}}
\newcommand{\rd}{{\mathbb R}^{2}}
\newcommand{\re}{{\mathbb R}}

\newcommand{\kn}{k \in {\mathbb Z}_+}

\newcommand{\gzkmpd}{\gamma_{0,k}^{(\beta)}(\mu + \delta)}
\newcommand{\gzkmmd}{\gamma_{0,k}^{(\beta)}(\mu - \delta)}

\newcommand{\nzkrd}{\nu_{0,k}(r_{\delta})}
\newcommand{\nzkrp}{\nu_{0,k}(r_{+})}
\newcommand{\nzkrm}{\nu_{0,k}(r_{-})}
\newcommand{\bp}{{\cal P}}
\newcommand{\bq}{{\cal Q}}
\newcommand{\bz}{{\cal Z}}

\makeatletter
\def\ps@wanda{%
   \renewcommand{\@oddhead}{\begin{minipage}{\textwidth}
        \footnotesize\itshape  Appeared in: {\rm Reviews in Mathematical Physics {\bf 14}, 1051--1072 (2002)}
         \end{minipage}                       }%
   \renewcommand{\@evenhead}{}%
   \renewcommand{\@oddfoot}{\hfil\footnotesize\textsf{\bfseries\thepage}\hfil}%
   \renewcommand{\@evenfoot}{}%
}%
\makeatother

\begin{document}
%
%------------------------------------------------------------------------------
  \thispagestyle{wanda} % only for the first page
%------------------------------------------------------------------------------
\vspace*{0.2cm}
\begin{center}
{\large\bf
  QUASI-CLASSICAL VERSUS NON-CLASSICAL \\ SPECTRAL ASYMPTOTICS
  FOR MAGNETIC SCHR\"ODINGER OPERATORS
  WITH DECREASING ELECTRIC POTENTIALS\\
}\end{center}
\vspace*{0.3cm}
\begin{center}
{GEORGI D. RAIKOV}\\[3.5pt] 
{\it  Departamento de Matem\'atica, 
      Universidad de Chile\\
      Las Palmeras 3425, Casilla 653, Santiago, Chile\\
      graykov@uchile.cl}\\
\vspace*{0.5cm}
{SIMONE WARZEL}\\[3.5pt] 
{\it  Institut f\"ur Theoretische Physik, 
      Universit\"at Erlangen-N\"urnberg\\
      Staudtstr. 7, D-91058 Erlangen, Germany\\
      warzel@physik.uni-erlangen.de}\\
%
%\vspace*{0.5cm}

%Received 25 December 2001\\
%Revised 28 June 2002
%\vskip2mm
%{Mathematical Physics Preprint archive: math-ph/0010013}\\
\end{center}
%\vskip4pt

%\vspace*{0.02cm}
%
%------------------------------------------------------------------------------
\begin{center}
\begin{minipage}[t]{129mm }
\noindent
We consider the Schr\"odinger operator $H(V)$ on 
$ L^2(\mathbb{R}^2 ) $ or $ L^2(\mathbb{R}^3 ) $, with
constant magnetic field and electric potential $V$ which typically decays 
at infinity exponentially fast or has a compact support. We 
investigate the asymptotic behaviour of the discrete spectrum of $H(V)$ 
near the boundary points of its essential spectrum. If the decay 
of $V$ is Gaussian or faster, this behaviour is non-classical in the
sense that  it is not described by the quasi-classical formulas known for the case 
where $V$ admits a power-like decay.

\vspace*{0.1cm}
{2000 Mathematics~Subject~Classification:~35P20,~47B35}\\
\end{minipage}
\end{center}

     \pagestyle{myheadings} % return to enumeration of pages
     \markboth{\hspace*{1cm} \it \small G.~D.~Raikov \& S.~Warzel \hfill}{%
      \hfill  \it \small Quasi-classical versus non-classical spectral asymptotics   \hspace*{1cm}}

\section{Introduction} 
Let $H(0) : = (-i\nabla - A)^2$ be the Schr\"odinger operator
with constant magnetic field of strength $b>0$, essentially
self-adjoint on $C_0^{\infty}(\re^d)$, $d=2,3$. The magnetic potential
$A$ is chosen in the form
$$
A({\bf x}) = \left\{
\begin{array} {l@{\quad\mbox{if}\quad}r}
\left(-\frac{by}{2}, \frac{bx}{2}\right) & d=2,\\[1ex]
\left(-\frac{by}{2}, \frac{bx}{2},0\right)  & d=3.
\end{array}
\right.
$$
In the two-dimensional case we identify the magnetic field with $\frac{\partial
A_2}{\partial x} - \frac{\partial A_1}{\partial y} = b$, while in the
three-dimensional case we identify it with ${\rm curl}\;A = (0,0,b)$. 
Moreover, if $d=2$, we write ${\bf x} = (x,y) \in \rd$, and if $d=3$,
we write ${\bf x} = (\xp, z)$ with  $\xp = (x,y) \in \rd$ and $z \in
\re$. Thus, in the latter case, $z$ is the variable along the magnetic
field, while $\xp$ are the variables on the plane perpendicular to it.  
Introducing the sequence of Landau levels $E_q: = (2q+1) b $, 
$q \in {\mathbb Z}_+ := \{0,1, \ldots \}$, we recall \cite{Foc,ahs} that
\begin{equation}\label{10}
\sigma(H(0)) = \sigma_{\rm ess}(H(0)) =  \left\{
\begin{array} {l@{\quad\mbox{if}\quad}r}
 \displaystyle\cup_{q=0}^{\infty}\{E_q\}  & d=2,\\[1ex]
\, [E_0,\infty) &  d=3. 
\end{array}
\right.
\end{equation}
Here $\sigma(H(0))$ denotes the spectrum of the operator
$H(0)$,
and $\sigma_{\rm ess}(H(0))$ denotes its essential spectrum. 

Let $V: \re^d \to \re$ be a measurable, non-negative function which decays at 
infinity in a suitable sense,  so that the operator
$V^{1/2}H(0)^{-1/2}$ is compact. By Weyl's theorem, 
$\sigma_{\rm ess}(H(0)) = \sigma_{\rm ess}(H(\pm V))$ where  
$H(\pm V) := H(0) \pm V$, and $ \pm V $ is the electric potential
of constant (positive or negative) sign.   

The aim of the article is to investigate the behaviour of the discrete
spectrum of the operator $ H(\pm V) $
near the boundary points of its essential spectrum. This behaviour has been extensively
studied in the literature in case where $V$ admits
power-like or slower decay at infinity (see \cite{so,t,r1,r3} or \cite[Chapters 11 and 12]{i}) 
and also in the special case where $ d= 3 $ and $ V $ is axially symmetric with respect to the magnetic field (see \cite{ahs,sol}). 
The novelty in the
present paper is that we consider $V$'s which decay
exponentially fast or have compact support and which at most asymptotically obey a certain symmetry. 
If $d=3$, this type of decay of $ V $ is supposed to take place in the directions perpendicular to the
magnetic field while the decay in the $z$-direction could be much more
general (see Theorems \ref{t22}--\ref{t22cs} below). If 
the decay of $V$ in the $(x,y)$-directions is Gaussian or super-Gaussian, we show that the 
discrete-spectrum behaviour of $H(\pm V)$ is not described by quasi-classical formulas known for the case 
of power-like decay. 

The results of the present paper have been announced in \cite{RW02}. 
After the initial submission of the paper, we became aware of the
preprint \cite{mr}. It deals with the eigenvalue asymptotics for the
Schr\"odinger and Dirac operators with full-rank magnetic fields, and
compactly supported electric potentials of fixed sign. In particular,
\cite{mr} extends our  Theorem \ref{t21cs}  to the case of {\em
  full-rank} magnetic fields in arbitrary even dimension. 
The methods of proof applied in \cite{mr} are variational ones similar to
those used in the present paper.

This paper is organized as follows. In Section 2 we formulate our 
main results. Section 3 is devoted to the analysis of the eigenvalue 
asymptotics for compact operators of Toeplitz type. Section 4 contains 
the proofs of the results concerning the two-dimensional case. 
Finally, the proofs of the results for the three-dimensional case
can be found in Section 5. 

\section{Formulation of Main Results}

\subsection{Basic notation}

In order to formulate our main results we need the following
notations. Let $T$ be a linear self-adjoint operator. Denote 
by ${\mathbb P}_I(T)$ the spectral projection of $T$ corresponding to
the open interval $I \subset \re$. Set 
\begin{align}
N(\lambda_1, \lambda_2; T) &:= {\rm rank}\;{\mathbb P}_{(\lambda_1,
\lambda_2)}(T), \quad \lambda_1, \lambda_2 \in \re, \quad 
\lambda_1 < \lambda_2, \notag \\
N(\lambda; T) & := {\rm rank}\;{\mathbb P}_{(-\infty,
\lambda)}(T), \quad \lambda \in \re.  \notag
\end{align}
If $T$ is compact, we will also use the notations 
\begin{equation}\label{20}
n_{\pm}(s;T): = {\rm rank}\;{\mathbb P}_{(s,\infty)}(\pm T), \quad
s>0.
\end{equation}
By $\|.\|$ we denote the usual operator norm, and by $\|.\|_{\rm HS}$ 
the Hilbert-Schmidt norm.

\subsection{Main results for two dimensions}
This subsection contains our main results related to the
two-dimensional case. 
 
\begin{theorem} \label{t21}
Let $V$ be bounded and non-negative on $\rd$. 
Assume that there exist 
two constants $ 0 < \mu < \infty $ and $ 0 < \beta < \infty $ such that  
\begin{equation}\label{21}
\lim_{|\bx| \to \infty} \frac{\ln V(\bx)}{|\bx|^{2 \beta}} = - \mu.
\end{equation}
Moreover, fix a Landau level $E_q$, $q \in \mathbb{Z}_+ $, and an energy 
$ E' \in (E_q, E_{q+1}) $. 
\begin{itemize}
\item[(i)] If ~$0 < \beta < 1$, then we have 
\begin{equation}\label{ch1}
\lim_{\ldz} \frac{N\big(E_q + E, E';H(V)\big)}{|\ln E|^{1/\beta}} =
\frac{b}{2\mu^{1/\beta}}. 
\end{equation}
\item[(ii)] If ~$\beta = 1$, then we have 
\begin{equation}\label{ch2}
\lim_{\ldz} \frac{N\big(E_q + E, E';H(V)\big)}{|\ln E|} = \frac{1}{\ln
(1+ 2\mu/b)}.     
\end{equation}
\item[(iii)] If ~$1 < \beta < \infty$, then we have 
\begin{equation}\label{ch3}
\lim_{\ldz} \frac{N\big(E_q + E, E';H(V)\big)}{(\ln |\ln E|)^{-1} |\ln
E|} = \frac{\beta}{\beta - 1}. 
\end{equation}
\end{itemize}\end{theorem}

The proof of Theorem \ref{t21} can be found in Subsection 4.2. It is 
evident from this proof that  Theorem \ref{t21} (iii) admits 
the following generalization as  the asymptotic coefficient in 
(\ref{ch3}) is independent of $\mu$. 

\begin{corollary} \label{f1} Let $V$ be bounded and non-negative on $\rd$. 
Assume  that there exist 
$ 0 < \mu_1 < \mu_2 < \infty$ and $ 1 < \beta < \infty $ such that  
$$
- \mu_2 \leq \liminf_{|\bx| \to \infty} \frac{\ln V(\bx)}{|\bx|^{2 \beta}},
\quad \limsup_{|\bx| \to \infty} \frac{\ln V(\bx)}{|\bx|^{2 \beta}} \leq - \mu_1.
$$
Then (\ref{ch3}) remains valid. 
\end{corollary} 

The last theorem of this subsection concerns the case where $V$ has a 
compact support. 

\begin{theorem} \label{t21cs}
Let  $V$ be bounded and non-negative on $\rd$. Assume  that the support of $V$ 
is compact, and there exists a constant $C_- > 0$ such that $V \geq C_-$ 
on an open non-empty subset of $\rd$. Moreover, let $q \in \mathbb{Z}_+ $ and $ E' \in (E_q, E_{q+1}) $. Then we have 
\begin{equation}\label{ch4}
\lim_{\ldz} \frac{N\big(E_q + E, E';H(V)\big)}{(\ln |\ln E|)^{-1} |\ln
E|} = 1. 
\end{equation}
\end{theorem}

The proof of Theorem \ref{t21cs} is contained in Subsection 4.3.

\begin{remark} Under the hypotheses of Theorems \ref{t21} or
\ref{t21cs} we have $V \in L^1(\rd) \cap  L^{\infty}(\rd)$. It is
well-known that this inclusion implies that the operator
$V^{1/2}(-\Delta + 1)^{-1/2}$ is compact. Hence, it follows from 
the diamagnetic inequality (see e.g. \cite{ahs}) that the 
operator $V^{1/2}H(0)^{-1/2}$ is compact as well.
\end{remark}

For further references, we introduce some additional notation
which allows us to unify (\ref{ch1})--(\ref{ch4}) into a single
formula. For  $ \kappa  \in (e , \infty)$ define the increasing
functions $a_\mu^{(\beta)}$ by    
\begin{equation}\label{Def:a}
  a_\mu^{(\beta)}(\kappa) := \left\{
    \begin{array}{c@{\qquad\mbox{\rm if}\qquad}c}
       \displaystyle \frac{b}{2} \left( \frac{\kappa}{\mu}\right)^{1/\beta}   & 0 < \beta < 1, \\[2ex]
       \displaystyle \frac{\kappa}{\ln \left(1 + 2 \mu/ b  \right)}           & \beta = 1, \\[2ex]
       \displaystyle \frac{\beta}{\beta - 1}\, \frac{\kappa}{\ln
    \kappa} & 1 < \beta < \infty, \\[2ex]
\displaystyle  \frac{\kappa}{\ln
    \kappa} & \beta = \infty.
      \end{array}
      \right.
\end{equation}
Then asymptotic relations (\ref{ch1})--(\ref{ch4}) can be re-written as 
\begin{equation}\label{21a}
  \lim_{E \downarrow 0} \,
  \frac{N\big(E_q + E, E';H(V)\big)}{a_\mu^{(\beta)}\big(| \ln E |
  \big)} = 1, \quad 0 < \beta \leq \infty. 
\end{equation}
\begin{remark} Whenever we refer to functions (\ref{Def:a}) with $1 <
\beta \leq \infty$, we will write $a^{(\beta)}(\kappa)$ 
instead of $ a_\mu^{(\beta)}(\kappa)$ because in this case they are
independent of $\mu$. 
\end{remark}

Let us discuss the results of Theorems \ref{t21} and
\ref{t21cs}. 
\begin{enumerate}
\item
      Asymptotic relation (\ref{21a}) describes the 
      behaviour of the  infinite sequence of discrete eigenvalues of the operator
      $H(V)$ accumulating to the Landau level $E_q$, $q \in \mathbb{Z}_+$, from the right.
      Analogous results hold if we consider the eigenvalues of $H(- V)$
      accumulating to $ E_q $ from the left. Namely, (\ref{21a}) remains valid
      if we replace $ N\big(E_q + E, E';H(V)\big)$ by 
      $ N\big(E'',E_q - E; H(-V)\big)$ with some $ E'' \in (E_{q-1},
      E_{q}) $ if $q>0$, or by $N(E_0-E;H(-V))$ if $q=0$.  
      
\item  Introduce the quasi-classical quantity      
      $$
       {\cal N}_{\rm cl}(E): = \frac{b}{2\pi} \left| \{{\bf x} \in \rd| V({\bf x}) > E\} \right|,
       \qquad E > 0, 
       $$ 
where $| \cdot |$ denotes the Lebesgue measure.
If $V \geq 0$ satisfies the asymptotics
    $V({\bf x}) =  |{\bf
x}|^{-\alpha} {\bf v}({\bf x}/|{\bf x}|)(1 + o(1))$ as $|{\bf x}| \to \infty$   
with some ${\bf v} \in C({\mathbb S}^1)$, ${\bf v} > 0$, and some $0 < \alpha <
    \infty $, then $\lim_{\ldz} E^{2/\alpha} {\cal N}_{\rm cl}(E) =
    \frac{b}{4\pi} \int_{{\mathbb S}^1} {\bf v}(s)^{2/\alpha} ds$, and
    it has been shown that 
       \begin{equation} \label{xyz}
         \lim_{E \downarrow 0 } \;\frac{N\big(E_q + E, E';
H(V)\big)}{{\cal N}_{\rm cl}(E)} = 1,   
       \end{equation}
assuming some  regularity of ${\cal N}_{\rm cl}(E)$ as
$ E \downarrow 0 $ (see \cite[Theorem 2.6]{r1}, \cite[Chapter 11]{i}). On the other hand, if $V$
       satisfies the assumptions of Theorem \ref{t21}, then
$$
\lim_{ E \downarrow 0 } \; \frac{{\cal N}_{\rm cl}(E)}{|\ln
       E|^{1/\beta}} = \frac{b}{2\mu^{1/\beta}}, \quad 0 <
       \beta < \infty, 
$$
and if $V$ satisfies the assumptions of Theorem \ref{t21cs}, then
$$
{\cal N}_{\rm cl}(E) = O(1), \quad \ldz. 
$$
Comparing (\ref{21a}) and (\ref{xyz}), we see that they are different
if and only if $ 1 \leq \beta \leq  \infty $. 
In case $ \beta = 1 $ the asymptotic orders of (\ref{21a}) and
(\ref{xyz}) coincide but their coefficients differ although they have
the same main asymptotic term  in the
       strong magnetic field regime $b \to \infty$. 
       In brief, asymptotic
       relation (\ref{21a}) is quasi-classical for potentials $ V$
whose decay is slower than Gaussian ($ 0 < \beta < 1 $), 
       and it is non-classical for potentials  
whose decay is faster than Gaussian ($ 1 < \beta \leq \infty $), while  
     the  Gaussian decay ($ \beta = 1 $) of $ V $ is the border-line
case.

A similar transition from quasi-classical to non-classical
  behaviour as a function of the decay of the single-site potential
  with Gaussian decay as the border-line case has been detected in
\cite{HuLeWa2}. 
There the leading low-energy
  fall-off of the integrated density of states of a charged quantum
  particle in $ \mathbb{R}^2 $ subject to a perpendicular constant
  magnetic field and repulsive impurities randomly distributed
  according to Poisson's law has been considered.

\item The assumptions of Theorems \ref{t21}--\ref{t21cs} that $V$ be 
bounded 
and non-negative are not quite essential. For example, both 
theorems remain valid if we consider potentials $|{\bf x}|^{-\alpha}
V({\bf x})$
where $0 < \alpha < 2$, and $V$ satisfies the hypotheses of Theorem
\ref{t21} or Theorem \ref{t21cs}.  
Similarly, Theorem \ref{t21} holds also in the case where $V$ is
allowed to change sign on  a compact subset  of $\rd$.

\item  Let $\pi(\lambda)$ be the number of primes
less than $\lambda > 0$. It is well-known that 
$$\lim_{\lambda \to \infty} \frac{\pi(\lambda)}{(\ln \lambda)^{-1}
\lambda} = 1$$ (see e.g. \cite[Section 1.8, Theorem 6]{haw}).  
Hence, (\ref{ch4}) can be re-written as 
$$\lim_{\ldz} \frac{N\big(E_q + E, E';H(V)\big)}{\pi (|\ln E|)} = 1.
$$ 

\end{enumerate}

\subsection{Main results for three dimensions} 
In this subsection we formulate our main results concerning
the case $d=3$. In this case we will analyze the behaviour of
$N(E_0-E;H(-V))$ as $\ldz$. 
In order to define properly the operator $H(-V)$ we need the following
lemma. 

\begin{lemma} \label{l21} Let $U \in L^1(\rd) \cap
L^{\infty}(\rd)$, and 
$v \in L^1(\re)$. Assume that 
$0 \leq V(\xp,z) \leq U(\xp) v(z)$, $\xp \in \rd$,  
$z \in \re \, $. Then the operator $V^{1/2} H(0)^{-1/2}$ is compact. 
\end{lemma}

The proof of the lemma is elementary. Nevertheless, for the reader's 
convenience we include it in Subsection 5.2. 

Denote by $H(-V)$ the self-adjoint operator generated in $L^2(\rt)$ by the quadratic
form 
$$
 \int_{\rt} \left\{|i\nabla u + Au|^2 - V|u|^2\right\} d\bx, \quad u
\in D(H(0)^{1/2}), 
$$
which is closed and lower bounded in $L^2(\rt)$ since  the operator
$V^{1/2} H(0)^{-1/2}$  
is compact  by Lemma \ref{l21}. 
\begin{theorem} \label{t22}
  Let $0 < \mu < \infty$ and $0 < \beta < \infty$. Assume that there exist a constant
      $C > 0$ and a
      function $v \in L^1(\re; (1 + |z|) dz)$, which does not vanish identically, such
      that
      $$
      0 \leq {V}(\bx) \leq C v(z), \quad \bx =(\xp,z) \in
      \mathbb{R}^3. 
$$ 
Moreover, suppose  that for every $\delta > 0$ there exist a constant 
      $r_\delta>0$ and two non-negative functions $ v^\pm_\delta \in  L^1(\re; (1 + |z|) dz) $,
      which do not vanish identically,
      such that 
      $$
    e^{-\delta|\xp|^{2 \beta}} \, v^-_\delta(z) \leq e^{\mu |\xp|^{2 \beta}} \, {V}(\xp, z) \leq
    e^{\delta|\xp|^{2 \beta}} \, v^+_\delta(z)  
      $$
      for all $|\xp| \geq r_\delta$ and all $ z \in \re $. 
 Then we have
  \begin{equation}\label{22a}
    \lim_{\ldz}\frac{N\big(E_0 - E;H(-V)\big)}{a_\mu^{(\beta)}\big(| \ln \sqrt{E} |
      \big)} = 1.
  \end{equation}
\end{theorem}
The proof of Theorem \ref{t22} can be found in Subsection 5.4.\\

Our last theorem treats the case where the projection of the support
of $V$ onto the plane perpendicular to the magnetic field is compact. 
Denote by $\chi_{r, \xp'}: \rd \to \re$ the
characteristic function of the disk
$\big\{\xp \in \rd |\; |\xp -  \xp'| < r\big\}$ of radius $r>0$,
centered at $\xp' \in \rd$. 
If $\xp' = 0$, we will write $\chi_r$ instead of $\chi_{r,0}$.
\begin{theorem} \label{t22cs}
  Assume that there exist four constants $r_{\pm} > 0$, $\xp^{\pm} \in \rd$, and two non-negative functions
  $ v^\pm  \in  L^1(\re; (1 + |z|)dz)$, which do not vanish identically, such that ${V}$ obeys the estimates
  $$
    \chi_{r_-, \xp^-}(\xp)\; v^-(z) \leq  {V}(\bx) \leq
    \chi_{r_+, \xp^+}(\xp) \; v^+(z), \quad \bx =(\xp,z) \in
      \mathbb{R}^3.
  $$
  Then we have
\begin{equation}\label{22acs}
  \lim_{\ldz}\frac{N\big(E_0 - E;H(-V)\big)}{a^{(\infty)}\big(| \ln \sqrt{E} |
  \big)} = 1.
\end{equation} 
\end{theorem}
The proof of Theorem \ref{t22cs} is contained in Subsection 5.5.\\

Let us discuss briefly the above results.
\begin{enumerate}
\item
In particular, Theorem \ref{t22} covers bounded negative
potentials $-V$ which decay at infinity exponentially fast, i.e. 
\begin{equation}\label{eq:Abfall}
\lim_{| \bx | \to \infty } \, \frac{\ln {V}(\bx)}{| \bx |^{2\beta} } = - \mu,
\end{equation}
with some $ 0<\beta <\infty $ and $ 0 < \mu < \infty $.

\item
Assume that $V \geq 0$ satisfies the asymptotics
    $V({\bf x}) =  |{\bf
x}|^{-\alpha} {\bf v}({\bf x}/|{\bf x}|) (1 + o(1))$ as $|{\bf x}| \to \infty$   
with some ${\bf v} \in C({\mathbb S}^2)$, ${\bf v} > 0$, and some $2 < \alpha <
\infty$. For $E > 0$ set 
$$
    \widetilde{{\cal N}}_{\rm cl}(E) := \frac{b}{2 \pi} \, \left| \left\{ {\xp} \in \rd \, \big|\;
      \int_{\mathbb{R}}  \, {V}(\xp,z) \; dz > 2 \sqrt{E} \, \right\} \right|.
$$
Under some supplementary regularity assumptions concerning the
behaviour of $\widetilde{{\cal N}}_{\rm cl}(E)$ as ${E\downarrow 0}$ we
have 
\begin{equation}\label{eq:class2}
    \lim_{E\downarrow 0} \, \frac{N\big(E_0 - E;H(-V)\big)}{\widetilde{{\cal
    N}}_{\rm cl}(E)} = 1 
  \end{equation}
(see \cite{so}, \cite[Theorem 1(ii)]{t},  \cite[Theorem 2.4(i)]{r1},
  \cite[Chapter 12]{i}). 
Theorem~\ref{t22} shows that (\ref{eq:class2}) remains valid if the decay of $ V $ is slower than Gaussian
  in the sense that 
  (\ref{eq:Abfall}) holds with $ 0 < \beta < 1 $. On the other hand,
  if this decay is Gaussian or 
faster in the sense that
  (\ref{eq:Abfall}) holds with $ \beta = 1 $ or $ 1 < \beta \leq \infty $,
  the asymptotics of $ N\big(E_0 - E;H(-V)\big) $
  as $ E \downarrow 0 $ differs from (\ref{eq:class2}).
\end{enumerate}
 
\section{Spectra of Auxiliary Operators of Toeplitz Type}

\subsection{Landau Hamiltonian and angular-momentum eigenstates}

Let $d=2$. In this case, by (\ref{10}) the spectrum of $H(0)$ 
consists of the eigenvalues $E_q$, $q \in
{\mathbb Z}_+$, which are of infinite multiplicity. Denote by $P_q$,  $q \in
{\mathbb Z}_+$, the spectral projection of $H(0)$ corresponding to the
eigenvalue $E_q$. Our next goal is to introduce convenient orthonormal
bases of the subspaces $P_q L^2(\rd)$. 
For ${\bf x} \in \rd$, $q \in {\mathbb Z}_+$, and $k \in {\mathbb Z}_+
- q: = \{-q,-q+1, \ldots\}$ we set 
\begin{equation}\label{EqAngular}
\varphi_{q,k}({\bf x}):=  \sqrt{\frac{q!}{(k+ q)!}} 
  \left[\sqrt{\frac{b}{2}}\, (x +  i y)\right]^k 
  {\rm L}^{(k)}_{q}\left(\frac{b \,|{\bf x}|^2}{2}\right) 
  \, \sqrt{\frac{b}{2\pi}} \,\exp\left(- \frac{b\, |{\bf x}|^2}{4}\right)
\end{equation}
where 
\begin{equation}\label{eq:Laguerre}
  {\rm L}^{(\alpha)}_{q}(\xi) 
    := \sum_{m=0}^{q} \binom{q+\alpha}{q-m} \frac{\left( - \xi \right)^m}{m!}, \qquad \xi \geq 0,
\end{equation}
are the generalized Laguerre polynomials (see e.g.\ \cite[Sec.~8.97]{GrRy}) which are defined in terms of the
binomial coefficients $ \binom{\alpha}{m} := \alpha (\alpha -1){\cdot} \dots {\cdot} (\alpha - m + 1) / m! $ 
if $ m \in \mathbb{Z}_+ \!\backslash \{0\} $, and $   \binom{\alpha}{0} := 1 $, for all $ \alpha \in \mathbb{R} $.   
It is well-known that the functions $\varphi_{q,k}$,  
$k \in \mathbb{Z}_+\! - q$, constitute an
orthonormal basis in the $q$th Landau-level 
eigenspace $P_q L^2(\rd)$, $q \in {\mathbb Z}_+$                         %
(see e.g. \cite{Foc,hlw}). In fact, $\varphi_{q,k}$ is also an
eigenfunction of the angular-momentum operator
$-i\left(x \,\partial/\partial y - y \, \partial/ \partial
x\right)$ with eigenvalue $k$. \\
 
For further references we establish some useful 
properties of the Laguerre polynomials ${\rm L}^{(\alpha)}_{q}\!$. We first recall \cite[Sec.~22.2.12]{abst} 
their orthogonality relation
    \begin{equation}\label{ortho}
    \int_0^\infty \! \xi^{\alpha} \, e^{-\xi} \,  {\rm L}^{(\alpha)}_{q}(\xi) \,  
     {\rm L}^{(\alpha)}_{q'}(\xi) \, d\xi = \frac{\Gamma(\alpha + q + 1)}{q!} \; \delta_{q,q'}
  \end{equation}
valid for all  $ q, q' \in \mathbb{Z}_+ $ and $ \alpha > -1 $. Here we have introduced Kronecker's delta $ \delta_{q,q'}$
and  Euler's gamma function $\Gamma(s): = \! \int_0^{\infty}\! t^{s-1} e^{-t} dt$, 
$s>0$, such that $\Gamma(k+1) = k!$ if $k \in
    \mathbb{Z}_+$, see e.g. \cite[Chapter 6]{abst}.  

\begin{lemma}\label{lemma:Laguerre}
   Let $ q \in \mathbb{Z}_+ $. Then  
  \begin{equation}\label{lemma:Laguerre1}
    \left| {\rm L}^{(k)}_{q}(\xi) \right| \leq (k+q)^q \; e^{\xi/(k+q)}
  \end{equation}
  holds for all $ \xi \geq 0 $ and all $ k \geq 1-q $. Moreover, one has the uniform convergence
  \begin{equation}\label{lemma:Laguerre2}
    \lim_{k\to \infty}  \, k^{-q}  \, {\rm L}^{(k)}_{q}(k\xi) = \frac{(1 - \xi)^q}{q!}
  \end{equation}
   for all $ 0 \leq \xi \leq 1 $.
\end{lemma}
\begin{remark} 
  An immediate consequence of (\ref{lemma:Laguerre2}) is the following lower bound on the pre-limit expression
  \begin{equation}\label{lemma:Laguerre22}
     k^{-q}  \, {\rm L}^{(k)}_{q}(k\xi) \geq \frac{(1-\xi_{0})^q}{2 \, q!} 
  \end{equation}
  which is valid for all $ 0 \leq \xi \leq \xi_{0} < 1 $ and sufficiently large $ k $.
\end{remark} 

\begin{proof}[\bf Proof of Lemma~\ref{lemma:Laguerre}]
  The rough upper bound (\ref{lemma:Laguerre1}) is taken from
  \cite[Eq.~(42)]{hlw}. 
  For a proof of (\ref{lemma:Laguerre2}) we use (\ref{eq:Laguerre})
  to obtain
  \begin{equation}\label{Laguerre2}
      k^{-q} \, {\rm L}^{(k)}_{q}(k\xi) = \sum_{m=0}^q    \, k^{m-q} \,
     \binom{q+k}{q-m} \,
     \frac{(-\xi)^m}{m!}.
  \end{equation}
  Asymptotic relation \cite[Eq.~6.1.46]{abst} entails 
  \begin{equation}\label{Eq:AbSteg}
    \lim_{k \to \infty} k^{m-q} \, \frac{\Gamma(k+q)}{\Gamma(k+m)} = 1.
  \end{equation}
  The r.h.s. of (\ref{Laguerre2}) thus converges (uniformly on $[0,1] $) 
  towards
  $
     \sum_{m=0}^q \binom{q}{m} (-\xi)^m/q! = (1 - \xi)^q/q!
  $
  by the binomial formula.
\end{proof}

For ${\bf x}, {\bf x}' \in \rd$ denote by $K_q({\bf x}, {\bf x}'): =
\sum_{k=-q}^{\infty} \varphi_{q,k}({\bf
x})\overline{\varphi_{q,k}({\bf x}')}$ the integral kernel of the 
projection $P_q$, $q \in \mathbb{Z}_+$. It is well-known that  
\begin{equation}\label{ch7}
K_q({\bf x}, {\bf x}') = \frac{b}{2\pi} {\rm
L}^{(0)}_{q}\left(\frac{b \,|{\bf x}-{\bf x}'|^2}{2}\right)
\exp{\left(-\frac{b}{4}\left(|{\bf x}-{\bf x}'|^2 + 2i(x'y -
xy')\right)\right)}
\end{equation}
(see e.g. \cite{hlw}). Note that we have 
\begin{equation}\label{ch8}
K_q({\bf x}, {\bf x}) = \frac{b}{2\pi}, \quad {\bf x} \in \rd, \quad q \in \mathbb{Z}_+. 
\end{equation}

\subsection{Compact operators of Toeplitz type}

In this subsection we investigate the eigenvalue
asymptotics of auxiliary compact operators of Toeplitz type $P_q \, F
P_q$ where $q \in \mathbb{Z}_+$ and $F$ is the multiplier by a
real-valued function. The
results obtained here  
will be essentially employed in the proofs of Theorems
\ref{t21}--\ref{t22cs}.

First of all, note that $P_q \, F P_q = e^{2(2q+1)bt} P_q \, e^{-t H(0) } F
e^{- t H(0) } P_q $,
$t>0$, $q \in \mathbb{Z}_+$. Hence, the diamagnetic
inequality implies that $P_q F P_q$ is compact if the operator
$|F|^{1/2}e^{\Delta t}$ (or, equivalently, $e^{\Delta t} |F|^{1/2}$) is
compact for some $t>0$ (see \cite[Theorems 2.2, 2.3]{ahs}).  In particular, the following
lemma holds.
\begin{lemma}\label{lcorrect}
{\rm ({\bf \cite[Lemma 5.1]{r1}})~}
  Let $ F $ be real-valued and $ F \in L^p(\rd)$ for some $p \geq 1$. 
Then the operator $ P_q F P_q $, $q \in \mathbb{Z}_+$, is self-adjoint and compact. 
\end{lemma}
\begin{lemma}\label{lemma:isotrop}
  Let $ F: \mathbb{R}^2 \to \mathbb{R} $  satisfy the conditions of 
  Lemma \ref{lcorrect}. Suppose in addition that $F$ is 
  radially symmetric with respect to the origin, and bounded.  Then the eigenvalues of the operator $ P_q F
  P_q $ with domain $ P_q L^2(\mathbb{R}^2) $, $q \in \mathbb{Z}_+$, are given by
\begin{equation}  
   \langle F \varphi_{q,k} \, , \, \varphi_{q,k} \rangle 
   = \frac{ q!}{(k+q)!} \int_0^\infty \! F\big( (
   \sqrt{2\xi/b} , 0 ) \big) \, e^{-\xi} \, \xi^k \, {\rm
   L}^{(k)}_{q}\left( \xi \right)^2 \, d \xi, \qquad k \in \mathbb{Z}_+\! -
   q, 
\end{equation}
where $\langle \cdot,\cdot\rangle$ denotes the scalar product in $L^2(\rd)$.  
\end{lemma}
\begin{proof}
  It suffices to take into account  
  (\ref{EqAngular}) 
  and the radial symmetry of $F$.
\end{proof}
\begin{remark} Evidently, Lemma \ref{lemma:isotrop} is valid under more
general assumptions. In particular, the boundedness condition is
unnecessarily restrictive. However, we state the lemma in a simple
form which is sufficient for our purposes.
\end{remark}

\subsection{ Two examples of explicit eigenvalue asymptotics}

For ${\bf x} \in \rd$ set
$G_\mu^{(\beta)}({\bf x}) := \exp\left( - \mu | {\bf x} |^{2\beta} \right) $ 
where $ 0 < \mu < \infty $ and $ 0 < \beta < \infty $.
According to Lemma~\ref{lemma:isotrop} the eigenvalues of 
$ P_q G_\mu^{(\beta)} P_q $ on $ P_q L^2(\mathbb{R}^2) $ are given by
\begin{equation}\label{Def:gamma}
  \gamma_{q,k}^{(\beta)}(\mu) := \big\langle G_\mu^{(\beta)} \varphi_{q,k} \, , \, \varphi_{q,k} \big\rangle  , \qquad
  k \in \mathbb{Z}_+ -q.
\end{equation}
Let $ \big(a_\mu^{(\beta)}\big)^{-1} $ denote the inverse function
of $ a_\mu^{(\beta)} $ defined in (\ref{Def:a}). 
Evidently, 
\begin{equation}\label{ch5}
\big(a_\mu^{(\beta)}\big)^{-1}(k) = 
      \left\{
      \begin{array}{c@{\qquad\mbox{\rm if}\qquad}c}
       \displaystyle \mu \left( \frac{2k}{b}  \right)^\beta   & 0 < \beta < 1, \\[2ex]
       \displaystyle k \, \ln \left(1 + 2 \mu/b \right)       & \beta = 1. \\[1ex]
       \end{array}
      \right.
\end{equation}
Moreover, it is straightforward to verify that 
\begin{equation}\label{ch6}
 \lim_{k \to \infty}\frac{\big(a^{(\beta)}\big)^{-1}(k)}{k \ln k} =  \,
      \left\{
      \begin{array}{c@{\qquad\mbox{\rm if}\qquad}c}
       \displaystyle \frac{\beta - 1}{\beta}     & 1 < \beta < \infty, \\[2ex]
       \displaystyle   1     & \beta = \infty. \\[1ex]
       \end{array}
      \right.
\end{equation}      
The next proposition treats the asymptotics of $\gamma_{q,k}^{(\beta)}(\mu)$, $q \in
\mathbb{Z}_+$, as $ k \to \infty$. 
For $ q= 0 $ and $ 0 < \beta \leq 1/2 $ closely related asymptotic evaluations can be found in \cite[Appendix]{sol}.

\begin{proposition}\label{Prop:gamma}
Let $ q \in \mathbb{Z}_+ $, $ 0 < \mu < \infty $, and $ 0 < \beta <
\infty $. Then we have 
\begin{equation}\label{Eq:Asygamma}
  \lim_{k\to\infty} \, \frac{\ln \gamma_{q,k}^{(\beta)}(\mu)}{\big(a_\mu^{(\beta)}\big)^{-1}(k)} = - 1
\end{equation}

\end{proposition}

\begin{proof}
From \eqref{Def:gamma} and Lemma~\ref{lemma:isotrop} 
it follows that
$
   \gamma_{q,k}^{(\beta)}(\mu) =  \frac{ q! \, k!}{(k+q)!} \, \mathcal{J}^{(\beta)}\big( k ,\mu {(2/b)}^\beta\big) 
$
where we have introduced the notation 
\begin{equation}\label{Def:M}
   \mathcal{J}^{(\beta)}\big( k , \lambda) := \frac{1}{k!} \,
   \int_0^\infty \! 
           \xi^{k}\, e^{- \lambda \xi^\beta - \xi } \,  {\rm
   L}^{(k)}_{q}\left( \xi \right)^2 \, d\xi .
\end{equation}
Thanks to asymptotic relation (\ref{Eq:AbSteg})
it remains to study the asymptotic behaviour of $ \mathcal{J}^{(\beta)} $ for large values of its first argument.
For this purpose we distinguish three cases.\\

\noindent
{\bf Case \boldmath $ 0 < \beta < 1 $.} 
    The claim follows from (\ref{Eq:AbSteg}) and (\ref{ch5}) with $0 < \beta <
    1$, together with the asymptotic 
    relation 
    \begin{equation}\label{Eq:AsymM1}
      \lim_{k \to \infty} 
       \, \frac{\ln \mathcal{J}^{(\beta)}\big( k , \lambda)}{k^{\beta}} = - \lambda
    \end{equation}
    valid for $ \lambda > 0 $ in this case. 
    For a proof of (\ref{Eq:AsymM1}) we construct asymptotically coinciding
    lower and upper bounds. To obtain a lower bound we suppose $ k > -1 $. The orthogonality relation (\ref{ortho}) implies that 
    $\xi^{k} \, e^{-\xi} \, 
        {\rm L}^{(k)}_{q}\left( \xi \right)^2  q!/ (k+q)! \;  d\xi$ induces a probability measure on $ [0, \infty ] $ such that Jensen's 
    inequality \cite{LiLo} yields  
    \begin{equation}
      \mathcal{J}^{(\beta)}\big( k , \lambda) 
       \geq \,\frac{(k+q)!}{k! \, q! } \, 
          \exp\left\{ - \lambda \,  \frac{q!}{(k+q)!} 
        \int_0^\infty \! \xi^{k+\beta} \, e^{-\xi} \, 
        {\rm L}^{(k)}_{q}\left( \xi \right)^2 \, d\xi \right\}.
    \end{equation}
    We may now employ the combinatorial identity 
    $ {\rm L}^{(k)}_q(\xi) = \sum_{m=0}^{q} \binom{m - \beta -1}{m} \,
    {\rm L}^{(k+\beta)}_{q-m}(\xi) $ \cite[Eq.~8.974(2)]{GrRy}, 
    which implies that 
    \begin{align} 
      & \frac{q!}{(k+q)!} \int_0^\infty \! \xi^{k+\beta} \, e^{-\xi} \, 
        {\rm L}^{(k)}_{q}\left( \xi \right)^2 \, d\xi \notag \\
      &  = \sum_{m,l=0}^q \binom{m - \beta -1}{m} \, \binom{l - \beta -1}{l}\,
        \frac{q!}{(k+q)!} \, 
        \int_0^\infty \! \xi^{k+\beta} \, e^{-\xi} \, 
        {\rm L}^{(k+\beta)}_{q-m}(\xi) \, 
        {\rm L}^{(k+\beta)}_{q-l}(\xi)  \, \, d\xi \notag \\
      &  = \sum_{m=0}^q \binom{m - \beta -1}{m}^2 \, \frac{q!}{(q-m)!} 
        \frac{\Gamma(k+q-m + \beta+1)}{\Gamma(k+q+1)}. 
     \end{align}
     Here we have again used the orthogonality relation (\ref{ortho}) in the 
     last step. 
     Using (\ref{Eq:AbSteg}) this entails 
     $ \liminf_{k \to \infty}  k^{-\beta} 
     \ln  \mathcal{J}^{(\beta)}\big( k , \lambda) \geq - \lambda $.\\
     For the upper bound we suppose $ k + q > 2 $ and choose 
    $ \Xi_k $ as the (unique) maximum of the integrand in the r.h.s.\
    of the estimate
    \begin{equation}\label{Def:M1}
      \mathcal{J}^{(\beta)}\big( k , \lambda) \leq \frac{(k+q)^{2q}}{k!} \,
      \int_0^\infty \!
      \xi^{k}\, e^{- \lambda \xi^\beta - (1-2/(k+q))\xi } \, d\xi
    \end{equation}
     which was obtained by using (\ref{lemma:Laguerre1}).
     More precisely, we define $ \Xi_k $ as the (unique) solution of the equation
    $ \lambda \beta \, \Xi_k^\beta + ( 1-2/(k+q)) \, \Xi_k = k $. 
    Splitting the integration in (\ref{Def:M1}) into two parts with domain 
    of integration restricted to $ [ 0 ,\Xi_k ) $ and $ [ \Xi_k , \infty ) $,
    the two parts are estimated separately as follows. 
    Using monotonicity of the integrand on $ [ 0 ,\Xi_k ) $ 
    we obtain the bound 
    \begin{align}
     & \frac{1}{k!} \, \int_0^{\Xi_k} \! \xi^{k}
           e^{- \lambda \xi^\beta - (1-2/(k+q))\xi } \, d\xi  
        \leq \frac{ \Xi_k^{k+1} }{k!} \, 
             \exp\!\big[- \lambda \Xi_k^\beta - (1-2/(k+q)) \Xi_k \big]  \notag \\
      & \mkern100mu    = \Xi_k \, \frac{k^k}{k!} \exp\!\big[k \ln\left[\Xi_k/k\right]  - (1-2/(k+q))\Xi_k - \lambda \Xi_k^\beta \big] \notag \\
       & \mkern100mu    \leq \Xi_k \frac{k^k}{k!} \, e^{-k } \, \exp\!\big[- \lambda \Xi_k^\beta + 2 \, \Xi_k/(k+q)\big] 
        \label{Eq:BoundM1}
    \end{align}
     on the first part.
    For the last inequality we have used the fact that $ \ln \xi \leq \xi -1  $ for all $ \xi > 0 $. The second part is 
    bounded according to
    \begin{align}
      \frac{1}{k!} \, \int_{\Xi_k}^\infty \!  \xi^{k} \, 
      e^{- \lambda \xi^\beta - (1-2/(k+q)) \xi } \, d\xi
      & \leq \exp\!\big[- \lambda \Xi_k^\beta \big] \, \int_0^\infty \!
      \frac{\xi^{k}}{k!} \,  e^{-(1-2/(k+q)) \xi}  \, d\xi \notag \\
      & = (1-2/(k+q))^{-k-1} \, \exp\!\big[- \lambda \Xi_k^\beta \big].  
      \label{Eq:BoundM2} 
     \end{align}
     The sandwiching bounds $ 1 - \lambda \beta k^{\beta -1} \leq  \left(1 - 2/(k+q)\right) \Xi_k / k 
     \leq 1 $ imply $  \lim_{k \to \infty} \Xi_k / k $\hspace{0pt}$ = 1$. Using this in 
     (\ref{Eq:BoundM1}) and (\ref{Eq:BoundM2}), employing
     Stirling's asymptotic formula \cite[Eq.~6.1.37]{abst}
     \begin{equation}\label{Eq:Stirling}
       \lim_{k\to \infty} \frac{k^{k-1/2}}{\Gamma(k)} \,  e^{-k} 
       = (2\pi)^{-1/2}, 
     \end{equation} 
     and the fact that 
     $ \lim_{k \to \infty} \left( 1 + 2/k \right)^k = e^2 $,   
     we obtain  
     $ \limsup_{k \to \infty}  k^{-\beta} \ln  \mathcal{J}^{(\beta)}\big( k , \lambda)$\hspace{0pt}$ \leq - \lambda $. This concludes the proof
     of (\ref{Eq:AsymM1}).\\

\noindent
{\bf Case \boldmath $  \beta = 1 $.}
    An explicit calculation yields
    \begin{align}
       & \mathcal{J}^{(1)}(k, \lambda)  = \frac{1}{k!} \,
       \int_0^\infty \!
           \xi^{k}\, e^{- (1+ \lambda) \xi } \,  
           {\rm L}^{(k)}_{q}\left( \xi \right)^2 \, d\xi  \notag \\
        &\quad = \frac{1}{k!} \sum_{m,l=0}^q \binom{q+k}{q-m} \binom{q+k}{q-l} \, 
        \frac{(-1)^{m+l}}{m! \,  l!} \, \int_0^\infty \!
           \xi^{k+m+l}\, e^{- (1+ \lambda) \xi } \, d\xi \notag \\
        & \quad =\sum_{m,l=0}^q \binom{q+k}{q-m} \binom{q+k}{q-l} \, 
        \frac{(-1)^{m+l}}{m! \,  l!} \, \frac{(k+l+m)!}{k!} \, (1+\lambda)^{-k-m-l-1}.
    \end{align}
    Using (\ref{Eq:AbSteg}) and proceeding similarly as in the second part of the
    proof of Lemma~\ref{lemma:Laguerre} one shows that
    the r.h.s. is asymptotically equal to
    \begin{equation}
      (1+\lambda)^{-k-1} \, \frac{k^{2q}}{(q!)^2} \, 
      \Bigg[ \sum_{m=0}^q \binom{q}{m} \frac{(-1)^m}{(1+\lambda)^m} \Bigg]^2 
      = (1+\lambda)^{-k-2q- 1} \, \frac{ {(\lambda \, k )}^{2q}}{(q!)^2}
    \end{equation}
    which in turn implies that $ \lim_{k \to \infty} k^{-1} \ln  \mathcal{J}^{(\beta)}(k,\lambda) = - \ln(1 + \lambda) $.\\

\noindent
{\bf Case \boldmath $ 1 < \beta < \infty $.} 
    The claim follows from (\ref{Eq:AbSteg}) 
    and (\ref{ch6}) together with the asymptotic 
    relation 
    \begin{equation}\label{Eq:AsymM2}
      \lim_{k \to \infty} 
      \frac{\ln \mathcal{J}^{(\beta)}\big( k , \lambda\big)}{k \,\ln k} = - \frac{\beta - 1}{\beta}
    \end{equation}
    valid for $ \lambda > 0 $ in this case. 
    For a proof of (\ref{Eq:AsymM2}) we construct asymptotically coinciding 
    lower and upper bounds. The lower bound reads 
    \begin{align}
       \mathcal{J}^{(\beta)}\big( k , \lambda\big) 
          & \geq  e^{- \lambda k - k^{1/\beta} } \, \frac{1}{k!} \, 
          \int_0^{k^{1/\beta}} \! 
            \xi^{k} \, 
           {\rm L}^{(k)}_{q}\left( \xi \right)^2 \, d\xi \notag \\
          &  \geq e^{- \lambda k - k^{1/\beta} }\, \frac{k^{k+1}}{k!}\,
            \int_0^{k^{\frac{1-\beta}{\beta}}} \! 
            \xi^{k} \, 
           {\rm L}^{(k)}_{q}\left( k \xi \right)^2 \, d\xi \notag \\
         & \geq e^{- \lambda k - k^{1/\beta} }\, \frac{k^{k+1/\beta}}{(k+1)!} \, k^{k\frac{1-\beta}{\beta}} \, 
         \frac{k^{2q}}{4^{q+1} \, (q!)^2}.
         \label{Eq:LoM2}
    \end{align}
    Here the last inequality follows from 
    (\ref{lemma:Laguerre22}) with $\xi_0 = 1/2$, and is valid for sufficiently large $ k $ only. 
    Using Stirling's asymptotic formula (\ref{Eq:Stirling}) in 
    (\ref{Eq:LoM2}), we obtain 
    $ \liminf_{k \to \infty} \big(k \,\ln k\big)^{-1}  $\hspace{0pt}$
    \ln \mathcal{J}^{(\beta)}\big( k , \lambda\big) \geq \frac{1-\beta}{\beta}  $.\\
    For the upper bound we suppose $ k + q > 2 $ and use
    \eqref{lemma:Laguerre1} in order to estimate the integrand in 
    (\ref{Def:M1}) from above. Thus we obtain
    \begin{equation}
      \mathcal{J}^{(\beta)}\big( k , \lambda\big) \leq 
      \frac{(k+q)^{2q}}{k!} \, \int_0^\infty \!
                  \xi^{k} \, e^{- \lambda \xi^\beta } \, d\xi 
                   = \frac{(k+q)^{2q} }{\beta \, \lambda^{(k+1)/\beta} \, k!} \;
    \Gamma\left(\frac{k+1}{\beta}\right). \label{Eq:UpM2}
  \end{equation} 
  Stirling's
  formula (\ref{Eq:Stirling}) finally yields 
  $ \limsup_{k \to \infty} \big(k \,\ln k\big)^{-1} $\hspace{0pt}$ 
    \ln \mathcal{J}^{(\beta)}\big( k , \lambda\big) \leq \frac{1-\beta}{\beta} $.
\end{proof}
The last topic in this section is the derivation of an asymptotic
property of the eigenvalues 
\begin{equation}\label{Def:nu}
   \nu_{q,k}(r) := \langle \chi_r \, \varphi_{q,k} \, , \, \varphi_{q,k} \rangle  , \quad
   k \in \mathbb{Z}_+ -q, \quad q \in \mathbb{Z}_+, \quad r>0, 
\end{equation}
of the operator $ P_q \chi_r P_q $ (see Lemma \ref{lemma:isotrop}). 
\begin{proposition}\label{Prop:nu} 
Let $ q \in \mathbb{Z}_+ $ and $ r > 0 $. 
Then we have
\begin{equation}\label{Eq:asynu}
  \lim_{k \to \infty } \, \frac{ \ln \nu_{q,k}(r) }{k \, \ln k } = - 1.
\end{equation}
\end{proposition}
\begin{remark} 
It follows from (\ref{Eq:asynu}), (\ref{Eq:Asygamma}), (\ref{ch5}),
and (\ref{ch6}) with $\beta < \infty$, that
  \begin{equation} \label{kiki}
    \nu_{q,k}(r) = o\big(\gamma_{q,k}^{(\beta)}(\mu)\big), \qquad k \to \infty,
    %\lim_{k \to \infty } \, \frac{\nu_{q,k}(r)}{\gamma_{q,k}^{(\beta)}(\mu)}  = 0  
  \end{equation}
  for all $ 0 < \mu < \infty $ and $ 0 < \beta < \infty $.
\end{remark}
\begin{proof}[\bf Proof of Proposition~\ref{Prop:nu}]
  From Lemma~\ref{lemma:isotrop} it follows that
  \begin{equation}\label{Eq:Sumnu}
    \nu_{q,k}(r) 
    = \frac{q! }{(k+q)!} \,   \int_{0}^{b r^2/2} \xi^{k} \, e^{-\xi} \,  {\rm
   L}^{(k)}_{q}\left( \xi \right)^2 \, d\xi .
  \end{equation}
  In its turn, the integral in (\ref{Eq:Sumnu}) is estimated as follows
  \begin{align}
    \int_{0}^{b r^2/2} \xi^{k} \, e^{-\xi} \,  
    {\rm L}^{(k)}_{q}\left( \xi \right)^2 \, d\xi 
    & \geq e^{- b r^2/2} k^{k+1} \int_{0}^{b r^2/(2k)} \! \xi^k \, {\rm L}^{(k)}_{q}\left( k \xi \right)^2\, d\xi \notag \\
    & \geq e^{- b r^2/2} \frac{k^{k+1}}{k+1} \, \left(\frac{b r^2}{2k}\right)^{k+1} \frac{k^{2q}}{4^{q+1} \, (q!)^2 }.
  \end{align}
   Here the last inequality  again is implied by  
    (\ref{lemma:Laguerre22}), and is valid for sufficiently large $ k $.  Moreover, we may use (\ref{lemma:Laguerre1}) to estimate
   \begin{equation}
     \int_{0}^{b r^2/2} \xi^{k} \, e^{-\xi} \,  
    {\rm L}^{(k)}_{q}\left( \xi \right)^2 \, d\xi  \leq
    (k+q)^{2q} \int_{0}^{b r^2/2} \xi^{k} \, e^{-(1-2/(k+q))\xi} 
     \leq \frac{(k+q)^{2q}}{k+1} \left(\frac{b r^2}{2} \right)^{k+1} 
  \end{equation}
  for all $ k + q \geq 2 $.
  The claim again follows with the help of  
  Stirling's formula (\ref{Eq:Stirling}).
\end{proof}
%%%%%%%%%%%%%%%%%%%%%%%%%%%%%%%%%%%%%%%%%%%%%%%%%%

\section{Proof of the Main Results for Two Dimensions} 

\subsection{Reduction to a single Landau-level eigenspace}

In this
subsection we establish  asymptotic estimates of 
$ N\big(E_q + E, E';H(V)\big)$ as $\ldz$, 
which play a crucial role in the proof of Theorems
\ref{t21} and \ref{t21cs}. 
For this purpose, we recall
in the following lemma a suitable version of the well-known
Weyl inequalities for the eigenvalues of self-adjoint compact
operators. 
\begin{lemma} \label{l41} {\rm ({\bf \cite[Section 9.2, Theorem 9]{bs}})~}
Let $T_1$ and   $T_2$ be linear self-adjoint compact operators on a
Hilbert space. Then for each $s > 0$ and $\varepsilon  \in (0,1)$ we have 
\begin{multline}\label{30}
n_{\pm}(s(1+\varepsilon); T_1) - n_{\mp}(s\varepsilon ; T_2) \leq 
n_{\pm}(s; T_1 + T_2) \\ \leq 
n_{\pm}(s(1-\varepsilon); T_1) + n_{\pm}(s\varepsilon; T_2), \quad   
\end{multline}
the counting functions $n_{\pm}$ being defined in (\ref{20}). 
\end{lemma}

\begin{proposition}\label{p41}  
Let $E' \in (E_q, E_{q+1})$, $q \in \mathbb{Z}_+$ .  
Assume that $V$ satisfies the hypotheses of Theorem \ref{t21} or
Theorem \ref{t21cs}. Then for every $\varepsilon \in 
  (0,1)$ we have 
\begin{align}\label{ch8a}
    n_+\big(E; (1-\varepsilon) P_q V P_q \big) +  
    O(1)  & \leq \,  
    N\big(E_q + E, E';H(V)\big)   \notag \\
& \leq  n_+\big(E; (1+\varepsilon) P_q V P_q\big) + O(1), \quad \ldz.   
\end{align}
\end{proposition}
\begin{proof}
First of all, note that under the hypotheses of Theorems
\ref{t21}--\ref{t21cs}, $V$ satisfies the assumptions of Lemma
\ref{lcorrect}, so that the operator $P_q V P_q$ is compact.

Next, the generalized Birman-Schwinger principle  (see
e.g. \cite[Theorem 1.3]{adr}) entails 
\begin{align}
 & N\big(E_q + E, E';H(V)\big)  \notag \\
 & \qquad \quad =  n_+\big(1; V^{1/2}(E_q + E - H(0))^{-1}V^{1/2}\big)  \notag  \\
  & \qquad \qquad \quad - n_+\big(1; V^{1/2}(E' - H(0))^{-1}V^{1/2}\big) -  {\rm dim \; Ker}\;
(H(V) - E'). \label{ch9}
\end{align}
 Since the operator $V^{1/2}H(0)^{-1/2}$ is compact, the last two terms at
the r.h.s.\ of (\ref{ch9}), which are independent of $E$, are 
finite.\\
Fix $\varepsilon \in (0,1)$ and set $Q_q: = {\rm Id} - P_q$. Applying  
(\ref{30}) with $T_1:= V^{1/2}(E_q + E - H(0))^{-1} P_q V^{1/2}$ 
and $T_2:= V^{1/2}(E_q + E - H(0))^{-1} Q_q V^{1/2}$, we obtain  
\begin{multline}
n_+\big(1; V^{1/2}(E_q + E - H(0))^{-1}V^{1/2}\big) \\ \geq  
n_+\big(1/(1-\varepsilon); V^{1/2}(E_q + E - H(0))^{-1}P_q
V^{1/2}\big)  \\ 
- n_-\big(\varepsilon/(1-\varepsilon); V^{1/2}(E_q + E - H(0))^{-1}
Q_qV^{1/2}\big), \label{ch10} 
\end{multline}
\begin{multline}
n_+\big(1; V^{1/2}(E_q + E - H(0))^{-1}V^{1/2}\big) \\ \leq 
n_+\big(1/(1+\varepsilon); V^{1/2}(E_q + E - H(0))^{-1}P_q
V^{1/2}\big)  \\
+ n_+\big(\varepsilon/(1+\varepsilon); V^{1/2}(E_q + E - H(0))^{-1} Q_q
V^{1/2}\big).  \label{ch11}
\end{multline} 
Next, we deal with 
the first terms on the r.h.s.\ of
(\ref{ch10}) and (\ref{ch11}). 
Since the non-zero singular numbers of the compact operators 
$P_q  V^{1/2}$ and $V^{1/2} P_q$ coincide, we get  
\begin{align}
& n_+\big(1/(1 \pm \varepsilon); V^{1/2}(E_q + E - H(0))^{-1}P_q
V^{1/2}\big) \notag \\ & \qquad = 
n_+\big(E; (1 \pm \varepsilon)V^{1/2}P_qV^{1/2}\big) \notag \\
& \qquad = 
n_+\big(E; (1 \pm \varepsilon) P_q V P_q\big). \label{ch12}
\end{align}
Further, we estimate the second terms on the r.h.s.\
of (\ref{ch10}) and (\ref{ch11}).  The operator inequality 
\begin{align}
|E_q + E - H(0)|^{-1} Q_q  & = \sum_{\substack{\; l \in \mathbb{Z}_+ \\ l \neq
    q}} |E_q + E - E_l|^{-1} P_l \notag \\
                            & \leq 
C_q \sum_{\; l \in \mathbb{Z}_+ } E_l^{-1}  P_l = C_q  \, H(0)^{-1},
\end{align}
valid for $E \in (0, E' - E_q)$, $E' \in (E_q,E_{q+1})$, and 
$C_q: = E_{q+1}/(E_{q+1} - E')$, implies
\begin{align}\label{ch14}
  & n_\pm\Big(\varepsilon/(1\pm\varepsilon); {V}^{1/2}(E_q + E - H(0))^{-1} Q_q \, {V}^{1/2}\Big) \notag \\
  & \qquad \quad \leq
  n_+\Big(\varepsilon/(1\pm\varepsilon); C_q \, {V}^{1/2}H(0)^{-1} {V}^{1/2}\Big).
\end{align}
Since the operator ${V}^{1/2}H(0)^{-1/2}$ is compact, the quantity on the
r.h.s.\ of \eqref{ch14}, which is independent of $E$,  
is finite for each $\varepsilon \in (0,1)$. 
Putting together (\ref{ch9})--(\ref{ch14}), we obtain (\ref{ch8a}).
\end{proof}  

\subsection{Proof of Theorem  \ref{t21}}
Pick $\delta \in (0,\mu)$. From~(\ref{21}) we conclude that there exist $r_\delta > 0$ such that
$ G_{\mu + \delta}^{(\beta)}({\bf x}) \leq V({\bf x}) 
  \leq G_{\mu - \delta}^{(\beta)}({\bf x})
$
for all $ {\bf x} \in \mathbb{R}^2 $ which satisfy $ |{\bf x}| > r_\delta $. Hence, we have 
\begin{equation}\label{4xxx}
  G_{\mu + \delta}^{(\beta)}({\bf x}) - M \chi_{r_\delta}({\bf x})
  \leq V({\bf x}) \leq G_{\mu - \delta}^{(\beta)}({\bf x}) + M  \,
  \chi_{r_\delta}({\bf x}), \quad {\bf x} \in \rd, 
\end{equation}
with $M: = \max\left\{1, \sup_{{\bf x} \in \rd} V({\bf x})\right\}$ as 
$\sup_{{\bf x} \in \rd}  G_{\lambda}^{(\beta)}({\bf x}) = 1$ for each
$\lambda \in (0,\infty)$, $\beta \in (0,\infty)$. 
Let us pick $\varepsilon > 0$. According to Proposition~\ref{p41} and
(\ref{4xxx}) 
  we have  
  \begin{align}
    N\big(E_q + E, E';H(V)\big) 
    & \geq n_+\big(E; (1-\varepsilon) P_q V P_q \big) + O(1),   \notag \\
    & \geq n_+\Big(E; (1-\varepsilon) 
    P_q \big[ G_{\mu + \delta}^{(\beta)} - M \chi_{r_\delta}  \big] P_q \Big) 
    + O(1), \quad \ldz,   \label{Eq:Mono1} 
  \end{align}
 \begin{align}
    N\big(E_q + E, E';H(V)\big)
    & \leq n_+\big(E; (1+\varepsilon) P_q V P_q \big) + O(1)  \notag \\
    & \leq n_+\Big(E; (1+\varepsilon) 
    P_q \big[ G_{\mu - \delta}^{(\beta)} + M  \chi_{r_\delta}  \big] P_q \Big) 
    + O(1), \quad \ldz.  \label{Eq:Mono2}
  \end{align}
  Since $ G_{\mu \pm \delta}^{(\beta)} \mp  M \chi_{r_\delta} $ is
  bounded and radially symmetric, 
  Lemma~\ref{lemma:isotrop} implies that the eigenvalues of 
  $ P_q \big[ G_{\mu \pm \delta}^{(\beta)} \mp  M \chi_{r_\delta}  \big] P_q $
  are given by 
  $ \gamma_{q,k}^{(\beta)}( \mu \pm \delta) \mp M \nu_{q,k}(r_\delta) $, $
  k \in \mathbb{Z}_+\! - q$, 
  (see (\ref{Def:gamma}) and (\ref{Def:nu})). Therefore, 
  \begin{multline}\label{eq:Anzahl1}
     n_+\Big(E; (1\mp\varepsilon) 
    P_q \big[ G_{\mu \pm \delta}^{(\beta)} \mp  M \chi_{r_\delta}  \big] P_q \Big) \\
     = \# \left\{ k \in \mathbb{Z}_+ - q \, \big| \, 
(1 \mp \varepsilon) \big[ \gamma_{q,k}^{(\beta)}( \mu \pm \delta) 
      \mp M \nu_{q,k}(r_\delta) \big]  > E \right\}, \quad 
  \end{multline}
Thanks to Proposition~\ref{Prop:gamma} and (\ref{kiki}), there
    exists some $ K_\varepsilon \in \mathbb{Z}_+-q $ 
such that 
  \begin{align}\label{eq:eigenvalues1}
    \gamma_{q,k}^{(\beta)}( \mu + \delta) - M \nu_{q,k}(r_\delta) 
    & \geq \left(1-\varepsilon\right) \gamma_{q,k}^{(\beta)}( \mu + \delta) \notag \\
    & \geq \left(1-\varepsilon\right) \, \exp\left[ - (1 +
    \varepsilon) \, 
\big( a_{\mu + \delta}^{(\beta)}\big)^{-1}(k) \right],   
  \end{align}
\begin{align}\label{eq:eigenvalues2}
    \gamma_{q,k}^{(\beta)}( \mu - \delta) + 
    M  \, \nu_{q,k}(r_\delta) 
    & \leq (1+\varepsilon) \, \gamma_{q,k}^{(\beta)}( \mu - \delta) \notag \\
    & \leq (1+\varepsilon) \, \exp\left[ - (1 - \varepsilon) \, 
\big( a_{\mu - \delta}^{(\beta)}\big)^{-1}(k) \right]  
  \end{align}  
  for all $ k \geq K_\varepsilon $. Using (\ref{Eq:Mono1})--(\ref{eq:eigenvalues2}), we thus conclude that
  \begin{align}\label{Eq:Down}
    & \liminf_{E \downarrow 0 } \, 
    \frac{N\big(E_q + E, E';H(V)\big)}{a_{\mu + \delta}^{(\beta)}\big(| \ln(E/\left(1-\varepsilon\right)^2)| / (1+\varepsilon)\big)} 
    \geq 1, \\ 
    \label{Eq:Up}
    & \limsup_{E \downarrow 0 } \, 
    \frac{N\big(E_q + E, E';H(V)\big)}{a_{\mu - \delta}^{(\beta)}\big(| \ln(E/\left(1+\varepsilon\right)^2)| / (1-\varepsilon)\big)} 
    \leq 1.
  \end{align}

Letting $ \varepsilon \downarrow 0 $ and afterwards $ \delta
\downarrow 0 $ in (\ref{Eq:Down}) and (\ref{Eq:Up}), and taking into
account  that
\begin{equation}
\lim_{\varepsilon \downarrow 0 } \, \lim_{\kappa \to \infty} \, 
  \frac{a_{\mu \pm \delta}^{(\beta)}\big( \kappa/ (1 \pm \varepsilon)\big)}{a_{\mu \pm \delta}^{(\beta)}\big( \kappa\big)}
  = 1, \quad
  \lim_{\delta \downarrow 0 } \, \lim_{\kappa \to \infty} \, 
  \frac{a_{\mu \pm \delta}^{(\beta)}\big( \kappa\big)}{a_{\mu}^{(\beta)}\big( \kappa\big)}
  = 1,
\end{equation}
we obtain (\ref{21a}) with $\beta < \infty$ which is equivalent to (\ref{ch1})--(\ref{ch3}). \qed

\subsection{Proof of Theorem \ref{t21cs}} 

Its hypotheses imply that there exist $C_{\pm} > 0$, $r_{\pm} > 0$, and
${\bf x}^{\pm} \in \rd$, such that 
\begin{equation}\label{fr1} 
C_-\;\chi_{r_-,{\bf x}^-}({\bf x}) \leq V({\bf x}) \leq 
C_+\;\chi_{r_+,{\bf x}^+}({\bf x}), \quad {\bf x} \in \rd. 
\end{equation}
Pick $\varepsilon \in (0,1)$. Combining (\ref{ch8a}), (\ref{fr1}), and
the minimax principle, we get 
\begin{align}\label{fr2}
N\big(E_q + E, E';H(V)\big) & \geq n_+\big(E; (1-\varepsilon)  C_-\;P_q
\;\chi_{r_-,{\bf x}^-}  P_q \big) + O(1), \quad \ldz, \\
\label{fr3}
N\big(E_q + E, E';H(V)\big) & \leq n_+\big(E; (1+\varepsilon)  C_+\;P_q
\;\chi_{r_+,{\bf x}^+}  P_q \big)
 + O(1), \quad \ldz. 
\end{align}
For ${\bf x}'=(x',y') \in \rd$ define the magnetic translation ${\cal
T}_{{\bf x}'}$ by 
$$
\left({\cal T}_{{\bf x}'}u\right)({\bf x}):=
\exp{\left\{i\frac{b}{2}(x'y-xy')\right\}}\,u({\bf x}-{\bf x}'), \quad {\bf x}
= (x,y) \in \rd. 
$$
The unitary operator ${\cal T}_{{\bf x}'}$ commutes with $H(0)$, and
hence with the projections $P_q$, $q \in \mathbb{Z}_+$ (see
e.g. \cite[Eq. 11]{hlw}). Therefore, 
\begin{equation}
P_q \chi_{r_{\pm},{\bf x}^{\pm}} P_q = P_q {\cal T}_{{\bf x}^{\pm}}
\; \chi_{r_{\pm}} {\cal T}_{{\bf x}^{\pm}}^* P_q = 
{\cal T}_{{\bf x}^{\pm}}\; P_q 
\; \chi_{r_{\pm}}  P_q \; {\cal T}_{{\bf x}^{\pm}}^*. 
\end{equation}
Hence, the operators $P_q \chi_{r_{\pm},{\bf x}^{\pm}} P_q$ and 
$P_q \chi_{r_{\pm}} P_q$ are unitarily equivalent, and we have 
\begin{align}
&  n_+\big(E; (1 \pm \varepsilon)  C_{\pm}\;P_q
\;\chi_{r_{\pm},{\bf x}^{\pm}}  P_q \big)  \notag \\
& \qquad \quad = 
 n_+\big(E; (1 \pm \varepsilon)  C_{\pm}\;P_q
\;\chi_{r_{\pm}}  P_q \big) \notag \\
& \qquad \quad = 
\# \big\{k \in \mathbb{Z}_+ - q \; | \; (1 \pm  \varepsilon) C_{\pm} 
\nu_{q,k}(r_{\pm}) > E\big\}  \notag \\
& \qquad\quad  = 
\# \big\{k \in \mathbb{Z}_+ - q \; | \;  
\ln \nu_{q,k}(r_{\pm}) + \ln((1 \pm  \varepsilon) C_{\pm}) > \ln
E\big\}. \label{fr4}
\end{align}
Taking into account (\ref{Eq:asynu}), we find that (\ref{fr4}) entails  
\begin{equation}\label{fr5}
\lim_{\ldz} \frac{ n_+\big(E; (1 \pm \varepsilon)  C_{\pm}\;P_q
\;\chi_{r_{\pm},{\bf x}^{\pm}}  P_q \big)}{(\ln |\ln E|)^{-1}|\ln E|} = 1.
\end{equation}
Putting together (\ref{fr2}), (\ref{fr3}), and (\ref{fr5}), we obtain (\ref{ch3}). \qed 
%%%%%%%%%%%%%%%%%%%%%%%%%%%%%%%%%%%%%%%%%%%%%%%%%%%%%%%%%%%%%%%%%%%%%%%%%

\section{Proof of Main Results for Three Dimensions}
\subsection{Auxiliary facts about Schr\"odinger operators in one
dimension} 
This subsection 
contains some well-known facts from the
spectral theory of one-dimensional Schr\"odinger operators.
  
Let $v \in L^1(\re)$ be real-valued and let 
$h(v)$ be the self-adjoint operator generated in $L^2(\re)$ by the
quadratic form 
$\int_{\re} \left\{|u'|^2 - v|u|^2\right\}\;dz$, $u \in W_2^1(\re)$. 
It is closed and lower bounded since the operator $|v|^{1/2}\big(h(0) +
1\big)^{-1/2}$ is Hilbert-Schmidt, and hence compact. 

\begin{lemma} \label{l51} {\rm ({\bf \cite[Subsections 2.4, 4.6]{b}, \cite{kl}})~}
Let $0 \leq v \in L^1(\re; (1 + |z|)dz)$, $g>0$. Assume that $ v $ does not vanish identically. 
Then we have
\begin{equation}\label{51}
1 \leq N(0; h(gv)) \leq g \int_{\re} |z| v(z) dz + 1.
\end{equation}
\end{lemma}
Note that if $0 < g \int_{\re} |z| v(z) dz < 1$, then by \eqref{51} the operator
$h(gv)$ has a unique, strictly negative eigenvalue denoted in the sequel
by $-{\cal E}(gv)$.

\begin{lemma} \label{l52} {\rm ({\bf \cite[Theorem 3.1]{bgs}, \cite{kl}, \cite{sol}})~}
Let the hypotheses of Lemma~\ref{l51} hold. Then  ${\cal E}(gv)$  obeys the
asymptotics
\begin{equation}\label{52}
\sqrt{{\cal E}(gv)} = \frac{g}{2} \int_{\re} v(z) dz \; (1 + o(1)), 
\quad g \downarrow 0.
\end{equation}
\end{lemma}

\subsection{Proof of Lemma \ref{l21}}
 Denote by $\bp_q: L^2(\rt) \to L^2(\rt)$, $q \in {\mathbb Z}_+$, the orthogonal projections
corresponding to the $q$th Landau level. In other words, 
$$
(\bp_q u)(\xp, z): = \int_{{\rd}} K_q(\xp, \xp') u(\xp',z) d\xp', \quad
(\xp, z) \in \rt, 
$$
where $K_q(\xp, \xp')$,
$\xp$, $\xp' \in \rd$, is the integral kernel of the orthogonal projection
$P_q:  L^2(\rd) \to L^2(\rd)$, introduced in (\ref{ch7}).

Let $N \geq 1$ and set $T:= V^{1/2}H(0)^{-1/2}$ and $T_N:= T
\sum_{q=0}^N \bp_q$. 

First, we show that $T_N$ is a Hilbert-Schmidt operator. To this end
we estimate 
$$\|T_N\|_{\rm HS} \leq \sum_{q=0}^N \|T \bp_q\|_{\rm HS}.$$ 
Further, taking into account (\ref{ch7})--(\ref{ch8}), we find that 
\begin{equation}
\|T \bp_q\|_{\rm HS}^2 = \frac{b}{(2\pi)^2} \int_{\rt} V({\bf x}) d{\bf x}
\int_{\re} \frac{d\zeta} {\zeta^2 + E_q}  \leq  \frac{b}{4\pi}
E_q^{-1/2} \, \|U\|_{L^1(\rd)} \, \|v\|_{L^1(\re)}.
\end{equation}
Therefore, $T_N$ is Hilbert-Schmidt, and hence compact.

Next, we show that $\lim_{N \to \infty}\|T -T_N\| = 0$. Evidently,  
\begin{equation}
\|T -T_N\| \leq  \|U\|^{1/2}_{L^{\infty}(\rd)}\;
\Big\| |v|^{1/2} \big( h(0) + E_{N+1}\big)^{-1/2}\Big\|.
\end{equation}
Since the operator $ |v|^{1/2} \big( h(0) + 1\big)^{-1/2}$ is compact in $ L^2(\re)$, we have  
$
\lim_{N \to \infty}\big\| |v|^{1/2} $ \vspace*{0pt} $\big( h(0) + E_{N+1}\big)^{-1/2}\big\| = 0  
$.
Consequently, the operator $T$ can be approximated in norm 
by the sequence of compact operators $T_N$. Hence, $T$ is a compact
operator itself. \qed

\subsection{Reduction to one dimension} 

In this subsection we prove a proposition which can be
regarded as the three-dimensional analogue of Proposition \ref{p41}. 
\begin{proposition} \label{p51}
Let $ {V} \geq 0 $. Suppose that there exist four non-negative functions
$ v^\pm \in L^1(\mathbb{R}) $ and
$ U^\pm \in L^1(\mathbb{R}^2) \cap L^\infty(\mathbb{R}^2) $ such that
\begin{equation}\label{eq:obun}
   U^-(\xp) \, v^-(z) \leq {V}(\bx) \leq  U^+(\xp) \, v^+(z), \quad \bx =
   (\xp, z ) \in \mathbb{R}^3. 
\end{equation}
Then for every $ \varepsilon > 0 $ we have 
\begin{align}
  \sum_{\kn} N\big(\!-\!E; h(\varkappa^-_k v^-)\big)
   & \leq  \, N\big(E_0-\!E; H(-{V}) \big)  \notag  \\
  &   \leq \sum_{\kn} N\big(\!-\!E; h((1+\varepsilon) \, \varkappa^+_k
   v^+)\big) + O(1), \quad \ldz. \label{emm}
\end{align}
Here $ h(v) $ is the operator 
defined at the beginning of Subsection 5.1, and $ \varkappa^\pm_k $, $
  \kn $, 
stand for the respective eigenvalues of
the compact operators $P_0 \, U^\pm \, P_0 $ on $P_0 L^2(\mathbb{R}^2) $.
\end{proposition}
\begin{proof}
Set $\bq_0: = {\rm Id} - \bp_0$ and denote by
$\bz_1(V)$ (respectively, by $\bz_2(V)$) the self-adjoint
operator generated in $\bp_0 L^2(\rt)$ (respectively, in $\bq_0 L^2(\rt)$) by the
closed, lower bounded quadratic form 
$\int_{\rt} \left\{|i\nabla u +  Au|^2 - V|u|^2\right\}
d\bx$  
defined for $u \in \bp_0 D(H(0)^{1/2})$ (resp., for  $u \in 
\bq_0 D(H(0)^{1/2}))$. Let $\varepsilon > 0$. 
Since $V \geq 0$, the minimax principle yields 
\begin{align}
N( E_0-E; \bz_1(V)) & \leq  N( E_0-E; H(-V)) \notag \\ 
\label{54}
& \leq N(E_0-E; \bz_1((1+\varepsilon)V))  \notag \\
& \qquad \quad + 
N(E_0-E; \bz_2((1+\varepsilon^{-1})V)).
\end{align}
It is easy to check that $\sigma_{\rm ess}(\bz_2((1+\varepsilon^{-1})V)) =
[E_1, \infty)$ for each $\varepsilon > 0$. Therefore,
\begin{equation}\label{55}
N(E_0-E; \bz_2((1+\varepsilon^{-1})V)) = O(1), \quad \ldz.
\end{equation}
Set $ V^{\pm}(\bx): = U^{\pm}(\xp) v^{\pm}(z)$, $\bx = (\xp, z)$.  
Then \eqref{eq:obun} implies   
\begin{align}\label{56}
& N\big( E_0-E; \bz_1(V)\big)  \geq N\big(E_0-E; \bz_1(V^-)\big), \\
\label{57}
& N\big(E_0-E; \bz_1((1+\varepsilon)V)\big)  \leq N\big(E_0-E; \bz_1((1+\varepsilon)V^+)\big). 
\end{align}
Obviously, $ \bz_1(V^-)$ is unitarily equivalent to the orthogonal
sum 
$\sum_{\kn} \oplus \big(h(\varkappa_k^- v^-) + E_0 \big) 
$, 
while   $\bz_1((1+\varepsilon)V^+)$ is unitarily equivalent to 
$\sum_{\kn} \oplus \big( h((1 + \varepsilon)\varkappa_k^+ v^+)+ E_0
\big)$.\\  
Thus the combination of (\ref{54})--(\ref{57}) yields 
(\ref{emm}). 
\end{proof}

\subsection{Proof of Theorem \ref{t22}} 
By the hypotheses of Theorem \ref{t22} we
may pick $ \delta \in (0,\mu) $ and choose $r_{\delta} > 0$ such that the assumptions of Proposition~\ref{p51}
are satisfied with
\begin{equation}
\begin{split}
 & U^\pm(\xp) = G_{\mu \mp \delta}^{(\beta)}(\xp) \pm {\cal M}\chi_{r_{\delta}}(\xp), \\
 & v^+(z) = v^+_\delta(z) + v(z), \;  v^-(z) = v^-_\delta(z),
\end{split}
\end{equation}
where, similarly to (\ref{4xxx}), ${\cal M}: = \max\{1,C\}$, and $C$ is the constant occurring in
the formulation of Theorem \ref{t22}. Accordingly, Lemma~\ref{lemma:isotrop} implies that
$ \varkappa^\pm_k = \gamma_{0,k}^{(\beta)}(\mu \mp \delta) \pm  {\cal M} \, \nzkrd $, $ k \in \mathbb{Z}_+ $.
Now pick $ \varepsilon \in (0,1) $ and choose $ K_{\varepsilon} $ such that
$k \geq K_{\varepsilon}$ entails the following  inequalities
\begin{equation}
\begin{split}
&\gzkmpd - {\cal M} \nzkrd  \geq (1 - \varepsilon) \, \gzkmpd, \\
& \gzkmmd + {\cal M} \nzkrd \leq (1 + \varepsilon)\, \gzkmmd, \\
&
(1 + \varepsilon)^2 \; \gamma_{0,k}^{(\beta)}(\mu \mp \delta)\int_{\re} \left| z \right| \, v^\pm(z) \, d z  < 1.
\end{split}
\end{equation}
Taking into account (\ref{51}) and 
Proposition~\ref{p51}, we get 
\begin{align}
  N(E_0 - E; H(-V)) & \geq \sum_{\kn} N\big(-E; h((\gzkmpd - {\cal M} \nzkrd)v^-)\big)
                               \notag \\
                  & \geq \# \left\{\kn, \; k \geq K_{\varepsilon} \, \big| \;
                    {\cal E}\big((1 - \varepsilon)\, \gzkmpd v^-\big)
                               > E \right\}.
                                \label{bel3}
\end{align}
Similarly, we have
\begin{align}
 & N(E_0 - E; H(-V)) \notag \\   
&\leq  \sum_{\kn} N\big(-E; h((1+\varepsilon)(\gzkmmd + {\cal M} \nzkrd)v^+\big) 
                               + O(1) \notag \\
& \leq \# \left\{\kn, \; k \geq K_{\varepsilon} \, \big| \;
                     {\cal E}\big((1 + \varepsilon)^2\, \gzkmmd
                               v^+\big) > E \right\} \notag \\
                             & \qquad \quad + O(1),  \; \ldz. \label{bel4}
\end{align}
The last inequality in (\ref{bel4}) results from splitting the series into two parts and using (\ref{51}) to verify that
the sum over $ k \in \{0, 1, \dots, K_{\varepsilon}-1 \} $ remains bounded as $ \ldz $. 
Utilizing (\ref{52}), choose $K'_{\varepsilon} \geq K_{\varepsilon} $
such that  
$k \geq K'_{\varepsilon}$
entails
\begin{align}\label{bel5}
\sqrt{{\cal E}\big((1 -\varepsilon)\, \gzkmpd v^-\big)} & \geq \frac{(1
-\varepsilon)^2}{2} \gzkmpd \int_{\re} v^-(z) \, dz, \\
\label{bel6}
\sqrt{{\cal E}((1 + \varepsilon)^2\gzkmmd v^+) } & \leq
\frac{(1+\varepsilon)^3}{2} \gzkmmd \int_{\re}  v^+(z) \, dz.  
\end{align}
Consequently, 
\begin{align}
  & \# \left\{\kn, \; k \geq K_{\varepsilon} \, |\;
   {\cal E}\big( (1 - \varepsilon) \, \gamma_{0,k}^{(\beta)}(\mu
   + \delta) v^{-}\big) > E \right\} 
 \notag \\ 
 & \quad \geq 
\# \left\{\kn, \; k \geq K_{\varepsilon}' \; \big|\; \frac{(1 - \varepsilon)^2}{2} \;
   \gamma_{0,k}^{(\beta)}(\mu + \delta)\big) \int_{\re} v^-(z) \;
   dz  >  \sqrt{E} \, \right\}, \label{bel5a} 
\end{align} 
\begin{align}
& \# \left\{\kn, \; k \geq K_{\varepsilon} \, |\;
   {\cal E}\big( (1 + \varepsilon)^2 \, \gamma_{0,k}^{(\beta)}(\mu
   - \delta) v^{+}\big) > E \right\} 
 \notag \\ 
 & \quad \leq \# \left\{\kn, k \; \geq K_{\varepsilon}' \; \big|\; \frac{(1 + \varepsilon)^3}{2} \;
   \gamma_{0,k}^{(\beta)}(\mu - \delta)\big) \int_{\re} v^+(z) \;
   dz  >  \sqrt{E} \, \right\} \notag \\
 & \qquad\quad+ O(1), \quad \ldz. \label{bel5b}
\end{align} 
Putting together (\ref{bel3})--(\ref{bel4}) and (\ref{bel5a})--(\ref{bel5b}), we
obtain the asymptotic estimates
\begin{align}\label{bel17}
 N\big(E_0-E; H(-V)\big)\!   & \geq  \,  \# \!\left\{\kn \, \big| \;
  \ln \gzkmpd  >  \ln{\sqrt{E}} + O(1)\right\}  + O(1),  \\
 N\big(E_0 - E; H(-V)\big) \! &  \leq  \,
\# \!\left\{\kn |\;  \ln \gzkmmd  >  \ln{\sqrt{E}} 
  + O(1) \right\}   + O(1),  \label{bel18}
\end{align}
valid as $\ldz$.
Using Proposition \ref{Prop:gamma} and proceeding as in the proof of Theorem~~\ref{t21},
we find that (\ref{bel17}) and (\ref{bel18})
imply  (\ref{22a}).  \qed 

\subsection{Proof of Theorem \ref{t22cs}} 
Finally, in this subsection we give a sketch of the
proof of Theorem \ref{t22cs} which is quite similar and only easier than
the proof of Theorem \ref{t22}. First of all, note that the assumptions of Proposition~\ref{p51}
are satisfied with
$
  U^\pm(\xp)  = \chi_{r_{\pm}, \xp^{\pm}}(\xp) $,
so that $ \varkappa^\pm_k = \nu_{0,k}(r_\pm) $ thanks to the unitary equivalence of the operators
$P_0  \chi_{r_{\pm}, \xp^{\pm}} P_0 $ and $P_0  \chi_{r_{\pm}} P_0 $
established in Subsection~4.3.
Proposition~\ref{p51} and Lemma~\ref{l51} then imply
the asymptotic estimates  
\begin{align}
& \# \left\{\kn \big| \,  \ln \nzkrm  >  \ln{\sqrt{E} + O(1)} \right\}
+ O(1) \notag \\
 & \quad \leq
N\big(E_0-E; H(-V)\big) \notag \\
 & \quad \leq
\# \left\{\kn \big|\,  \ln \nzkrp >  \ln{\sqrt{E} + O(1)} \right\}
+ O(1),   \label{bel9}
\end{align} 
which hold for $\ldz$, and are  analogous to (\ref{bel17}) and (\ref{bel18}).
Applying (\ref{Eq:asynu}) and (\ref{ch6}) with $\beta = \infty$, 
we conclude that (\ref{bel9})
implies (\ref{22acs}). \qed 

\section*{Acknowledgements}
 
The authors are very grateful to Professor Grigori Rozenblum
for indicating a gap  
in the proof of Propositions~\ref{Prop:gamma} and \ref{Prop:nu} 
for $ q \geq 1 $ in the first version of the paper. Acknowledgements are also
due to both referees whose remarks contributed to the improvement of
the article. 

A part of this work was done while G.~Raikov 
was visiting the Friedrich-Alexander Universit\"at Erlangen--N\"urnberg in
the summer of 2001 as a DAAD Research Fellow. The financial support
of DAAD and of the Chilean Science Foundation {\it Fondecyt} under Grants 1020737 and 7020737, is gratefully acknowledged.

It is a pleasure for G.~Raikov to express his gratitude to Professor Hajo Leschke
for his warm hospitality. 
Both authors thank him for
encouragement and several stimulating discussions.\\

%\vspace{0.5cm}
%{\bf Note Added in Proof:}~ After submission of the present paper, we learned of the work 
%{\em Eigenvalue asymptotics for even-dimensional perturbed Dirac and Schr\"odinger operators with
%constant magnetic field}, mp\_arc 02-140, by M. Melgaard and G. Rozenblum, which 
%contains a proof of
%Theorem~2.2 similar to the one in the present paper. Moreover, it 
%deals with its generalization of Theorem~2.2 to higher even dimensions and the case of the Dirac operator.\\  

\end{document}